\documentclass[aps,prb,twocolumn,floatfix,groupedaddress,longbibliography]
{revtex4-2}
\usepackage[T1]{fontenc}
\usepackage{amsmath} 
\usepackage{amssymb} 
\usepackage{amsfonts} 
\usepackage{bm} 
\usepackage{graphicx} 
\usepackage{hyperref} 
\usepackage{xcolor} 
\usepackage{float} 
\usepackage{subfigure} 
\usepackage{array} 

\begin{document}
\title{
A toy model for 2-dimensional spin-fluctuation-induced 
unconventional superconductivity
}
\author{Tu M. Cao}
\affiliation{Department of Physics and Astronomy, George Mason University, Fairfax, VA 22030}
\affiliation{Center for Quantum Science and Engineering, George Mason University, Fairfax, VA 22030}
\author{Igor I. Mazin}
\affiliation{Department of Physics and Astronomy, George Mason University, Fairfax, VA 22030}
\affiliation{Center for Quantum Science and Engineering, George Mason University, Fairfax, VA 22030}

\begin{abstract}
Superconductivity had been one of the most enigmatic phenomena in condensed matter physics, puzzling the best theorists for 45 years, since the original discovery by Kamerlingh-Onnes in 1911 till the final solution by Bardeen, Cooper and Schrieffer (BCS) in 1957. The original BCS proposal assumed the highest-symmetry form for the superconducting order parameter {$\Delta$},
namely, a constant, and a uniform pairing interaction due to attractive mediation of ionic vibration. 
While it was rather soon realized that generalizations onto k-dependent order parameters and anisotropic pairing interaction was straightforward,
only thirty years later, upon the discovery of high-temperature superconductivity in cuprates,  high-order angular dependence of \texorpdfstring{$\Delta$}{} and repulsive interaction, mediated by spin fluctuations or Coulomb repulsion brought such “unconventional'' into the spotlight. In 2008 yet another such system was discovered, and eventually the idea of repulsion-mediated unconventional superconductivity was generally accepted. Apart from the two specific systems mentioned above, a large number of various specific implementations of this idea have been proposed, and it is becoming increasingly clear that it is worth studying mathematically how unconventional superconductivity emerges, and with what properties, for a simple, but sufficiently general theoretical model. In our project, we study systematically unconventional superconductivity in an isotropic two-dimensional model system of electrons, subjected to repulsive interactions of a simple, but physically motivated form: a delta function peaked at a particular momentum (from 0 to twice the Fermi momentum), or Gaussian of varying widths. 
\end{abstract}

\maketitle

\section*{Introduction}
Even though theorists were taking random forays into unconventional superconductivity (``unconventional'' here is defined as superconductivity that is not due to phonons, and/or with a momentum-dependent order parameter\cite{Mazin_2024}), the real advent of this field was triggered by the discovery of the high-temperature superconductivity in cuprates\cite{5}. It took a decade to appreciate that the pairing symmetry there is d-wave, and that the likely pairing interaction is due to spin-fluctuation exchange\cite{6}. This concept has gained currency afterwards, especially when it was recognized that the newly-discovered Fe-based superconductors, even while qualitatively different and of different pairing symmetry, can also be described within the same paradigm. 

The idea is tantalizingly simple. As we recall in the next section, while charge fluctuations, such as phonons, always induce inter-electron attraction, and Coulomb interaction is always repulsive, spin fluctuations induce repulsion for singlet pairs, and attraction in triplet pairs (we are not discussing even more exotic odd-frequency superconductivity, where this rule is reversed, even though this concept has also been brought up within the same framework\cite{odd}). If, as it is usually the case, superconductivity develops upon suppression of an antiferromagnetic order, spin-fluctuation spectrum is peaked, in the momentum space, at the wave vector corresponding to this order. If the Fermi surface geometry is such that this vectors spans parts of the Fermi surface that, in a given pairing symmetry, have opposite signs, spin-fluctuation exchange will favor this particular pairing symmetry. In case of high-$T_c$ cuprates it happens to be the $x^2-y^2$ $d$-wave symmetry, in Fe-based material a sign-changing $s$-wave\cite{4}, etc. Similar geometrical arguments were historically applied for triplet pairing, such as $p$-wave\cite{Rice_1995} or $f$-wave\cite{Mazin2014}. 

Apart from some heavy-fermion superconductors, most unconventional-superconductivity candidates are 2D, which greatly simplifies the problem. Keeping in mind that real materials usually have complex Fermi surfaces (high-$T_c$ cuprates are an exception) and spin fluctuations often have a complicated spectrum, it is still useful to gather general insight into the interplay between the Fermi surface geometry and the shape of the spin fluctuation spectrum using simple models.

In this paper, we are using a minimal 2D model, consisting of a circular Fermi surface of a unit radius $k_F=1$ and and isotropic spin-fluctuation induced pairing interaction, $V_{\mathbf{k,k'}}=V(|\mathbf{k-k'}|)$, and considering both singlet states up to $l=4$ (i.e., $s$, $d$ and $g$) and triplet up to $l=5$ (i.e., $p$, $f$ and $h$). The high symmetry of the model affords a great simplification of the mathematics involved, without much loss of the essential physics.

\section*{General Theory}
\subsection*{Anisotropic Bardeen-Cooper-Schrieffer (BCS) Theory}
The BCS theory introduced the concept of a uniform order parameter \( \Delta \) and a constant, attractive pairing interaction \( g=V_{k, k'} \). The standard BCS equation then reads:
\begin{equation}
    \Delta = g \sum_{\mathbf{k}} \frac{\Delta}{2  
E_{\mathbf{k}}} \tanh\left(\frac{E_{\mathbf{k}}}{2T}\right)
   \approx g \sum_{\mathbf{k}} \frac{\Delta}{2 \epsilon_{\mathbf{k}}} 
   \tanh\left(\frac{\epsilon_{\mathbf{k}}}{2T}\right)
    \label{BCS}
\end{equation}
where $E_{\mathbf{k}} = \sqrt{\Delta^{2}+\epsilon_\mathbf{k}^{2}}$ is the excitation energy in the superconducting state, $\epsilon_\mathbf{k}$ is the normal-state one-electron energy, with the Fermi energy $E_F$ set to zero, and $g> 0$ 
is the attractive constant interaction; the second equality holds in the linear regime $T_c-T\ll T_c$, $\Delta\ll T_c$. Furthermore, the interaction is 
presumed to be non-zero only for $\epsilon_\mathbf{k}<T_D$, a cut-off frequency. In the assumed weak-coupling regime (not to be confused with the weak-coupling limit of the Eliashberg theory) the dimensionless coupling constant $\lambda=gN\ll 0$ (where $N$ is the density of states at the Fermi level). Going from integration over the momenta to integration over energies in Eq. \ref{BCS}, one obtains the linearized equation on $T_c$:
\begin{equation}
    \Delta = \lambda\int_0^{T_D}  \frac{\Delta}{2 \epsilon}\tanh\left(\frac{\epsilon}{2T}\right)d\epsilon.
    \label{BCS2}
\end{equation}

The (small) order parameter can be cancelled out and the remaining equation is easily solved in the $T_c\ll T_D$ limit to give
\[
T_c=1.13T_D \exp{(-1/ \lambda)}
\]
where 1.13  comes from the Euler $\gamma$ as $2e^\gamma/\pi$.

A straightforward generalization of the Bardeen-Cooper-Schrieffer (BCS) theory allows for the momentum dependence of both \(V(\mathbf{k}, \mathbf{k'})\) and \( \Delta(\mathbf{k})\). The gap equation then becomes:
\begin{eqnarray}
    \Delta_{\mathbf{k}} &=&\sum_{\mathbf{k'}} V_{\mathbf{k}, \mathbf{k'}} N(\mathbf{k'}) \Delta_{\mathbf{k'}} \log\left(\frac{1.13 T_D}{T_c}\right)\nonumber\\
    &=&\sum_{\mathbf{k'}} \lambda_{\mathbf{k}, \mathbf{k'}}\Delta_{\mathbf{k'}} \log\left(\frac{1.13 T_D}{T_c}\right)
    \label{eq:s}
\end{eqnarray}
where \( N(\mathbf{k'}) = \frac{1}{v_F(\mathbf{k'})} \) is the local density of states at the Fermi surface, with \( v_F(\mathbf{k'}) \) being the Fermi velocity. 

In the proximity of weak-coupling limit on an anisotropic Fermi surface, the order parameter equation can be expressed as an eigenvalue problem:
\begin{equation}
    \sum_{\mathbf{k'}} \lambda_{\mathbf{k}, \mathbf{k'}}\Delta_{\mathbf{k'}}= \frac{1}{\log{(1.13T_D/T_c)}}\Delta_{\mathbf{k}} \label{3}
\end{equation}

The largest eigenvalue $\lambda_{\max}$ of the matrix 
$\lambda_{\mathbf{k},\mathbf{k'}}$ thus gives the 
largest critical temperature $T_c$ at which a 
solution of the Eq. \ref{3} is possible, and the corresponding eigenvector $\Delta_{\mathbf{k}}$ gives us the corresponding distribution of the order parameter over the Fermi surface near $T_c$ (but not at zero temperature). Then
\begin{equation}
    T_c = 1.13 T_D \exp(-1/\lambda_{\max})
\end{equation}
in this formulation, \( \lambda_{\max} \) replaces the typical coupling constant \( \lambda \) in the conventional BCS theory. Note that, in principle, the order parameter $\Delta$ need not be real, but may have a complex phase. However it must satisfy (see the next section) the requirement that $\Delta_{\mathbf{k}} = \Delta_{-\mathbf{k}}$ (we are not considering non-centrosymmetric crystal lattices here), so any eigenvector that does not respect this condition, even if it yields the largest eigenvalue, should be discarded.

\subsection*{Generalization onto triplet pairing}
While the BCS theory assumes singlet pair with the opposite spins, a similar theory can be written for triplet pairs, where each pair has spin $S=1$\cite{1}. Since the pair is now a spin-1 object, its state has to be described by a spinor matrix, which, in turn, can be represented by a real-space axial vector. Furthermore, while in the singlet case the pair wave function satisfy the Pauli principle by virtue of its spin part, so that its spatial part  $\Delta_{\mathbf{k}} = \Delta_{-\mathbf{k}}$ is inversion-symmetric, the opposite is true for the triplet case, so the vector order parameter is antisymmetric: $\mathbf{d_{k}} =-\mathbf{d_{-k}}$.

One can now write BCS-like equations on this vector order parameter:
\begin{eqnarray}
    \mathbf{d}_{\mathbf{k}} &=&\sum_{\mathbf{k'}} V_{\mathbf{k}, \mathbf{k'}} N(\mathbf{k'}) \mathbf{d}_{\mathbf{k'}} \log\left(\frac{1.13 T_D}{T_c}\right)\nonumber\\
    &=&\sum_{\mathbf{k'}} \lambda_{\mathbf{k}, \mathbf{k'}}\mathbf{d}_{\mathbf{k'}} \log\left(\frac{1.13 T_D}{T_c}\right)\label{d}
\end{eqnarray}

Possible symmetries of the vector $\mathbf{d}$ are enumerated, for the three most common crystal symmetries, cubic, tetragonal and hexagonal, in the review Ref. \cite{1}. For the purpose of our minimal model, they can be greatly simplified, since, first, we only need to consider 2D representation, and, second, neglecting spin-orbit coupling essentially renders all triplet unitary states with the same angular momentum degenerate. For instance, for a tetragonal or hexagonal system Sigrist and Ueda\cite{1} list four unitary states, $\mathbf{d_\mathbf{k}}=const\cdot(k_x\mathbf{\hat{x}}\pm k_y\mathbf{\hat{y}})$, or $const\cdot(k_x\mathbf{\hat{y}}\pm k_y\mathbf{\hat{x}})$, which are all degenerate. Thus, it is enough to consider only $\mathbf{d_\mathbf{k}}=const\cdot(k_x\mathbf{\hat{x}}+ k_y\mathbf{\hat{y}})=const\cdot \mathbf{k}/k$. Correspondingly, triplet states with higher angular momenta that $p$ ($l=1$) can be, without a loss of generality, written as  $\mathbf{d_\mathbf{k}}=\Delta^T(\mathbf{k}) \mathbf{k}$, where, according to our model, $k=1$, and $\Delta^T$ is a scalar  {\em inversion-symmetric} function.

Substituting this form into Eq. \ref{d}, we get 
\begin{eqnarray}
      \Delta^T_\mathbf{k}{\mathbf{k}} &=&\sum_{\mathbf{k'}} \lambda_{\mathbf{k}, \mathbf{k'}}\Delta^T_\mathbf{k'}{\mathbf{k'}} \log\left(\frac{1.13 T_D}{T_c}\right)\\
      \Delta^T_\mathbf{k} &=&\sum_{\mathbf{k'}} \lambda_{\mathbf{k}, \mathbf{k'}}\Delta^T_\mathbf{k'}\mathbf{k}\cdot{\mathbf{k'}} \log\left(\frac{1.13 T_D}{T_c}\right)
      \label{d2}
\end{eqnarray}
which has the same form as for the singlet pairing, but replacing the $\lambda_{\mathbf{k}, \mathbf{k'}}$ matrix with $\lambda_{\mathbf{k}, \mathbf{k'}}(\mathbf{k}\cdot{\mathbf{k'}})$. Importantly, the interaction matrix $\lambda$ in the singlet case is, for the same spin fluctuation spectrum, three times larger, due to spin-rotational invariance\cite{1}. In order to keep the same notations for both case, we now replace Eq. \ref{eq:s} with the following:
\begin{eqnarray}
      \Delta^S_\mathbf{k} &=&3\sum_{\mathbf{k'}} \lambda_{\mathbf{k}, \mathbf{k'}}\Delta^S_\mathbf{k'} \log\left(\frac{1.13 T_D}{T_c}\right)
      \label{d2}
\end{eqnarray}
\section*{Simplified Model of Spin-Fluctuation-Induced Interaction}
When one includes all the aforementioned generalizations of the BCS theory, the phase diagram of the resulting superconducting state becomes rather complex. Some qualitative understanding can be gained from a simple toy model of a uniform 2D electron gas with an isotropic spin-fluctuation induced interaction.

Thus, we take the spin-fluctuation pairing interaction to be $V(\mathbf{k,k}^{\prime}) = V f(|\mathbf{k - k}^{\prime}|)$, which is presumed to have a peak at a momentum \( Q \). We will consider two models for $V$: first, a Dirac-\(\delta\) function, and, second, a Gaussian with a finite width $\kappa$. The Gaussian model is more realistic, but the $\delta$-function model allows for an analytical solution and serves as a limiting test case when $\kappa \rightarrow 0$. The form of the interaction is, respectively,

\begin{equation*}
\begin{aligned}
    f &= \delta(|\mathbf{k-k}^{\prime}|-Q)  \\
    f &= \frac{1}{\kappa\sqrt{\pi}}\exp\left[-\frac{(|\mathbf{k-k}^{\prime}|-Q)^{2}}{\kappa^{2}}\right]
\end{aligned}
\end{equation*}
Note that for \( Q = |\mathbf{k - k'}| \):
\begin{equation*}
    Q = \sqrt{\mathbf{k}^2 + \mathbf{k'}^2 - 2 k k' \cos(\tilde{\varphi})} = \sqrt{2 - 2 \cos(\tilde{\varphi})}
\end{equation*}
where \(\tilde{\varphi}\) is the angle between \(\mathbf{k}\) and \(\mathbf{k'}\), and \(|\mathbf{k}|\), \(|\mathbf{k'}|\) are normalized to 1. This implies that the parameter \( Q \) has values in the interval \([0, 2]\), with \( Q = 0 \) when \(\mathbf{k}\) and \(\mathbf{k'}\) overlap, and \( Q = 2 \) when they are opposite. We do not consider cases where the peak in the spin fluctuation spectrum is outside of the Fermi surface ($Q>2)$

\subsection*{Angle representation}
\begin{figure}[h]
    \centering \includegraphics[width=0.75\linewidth]{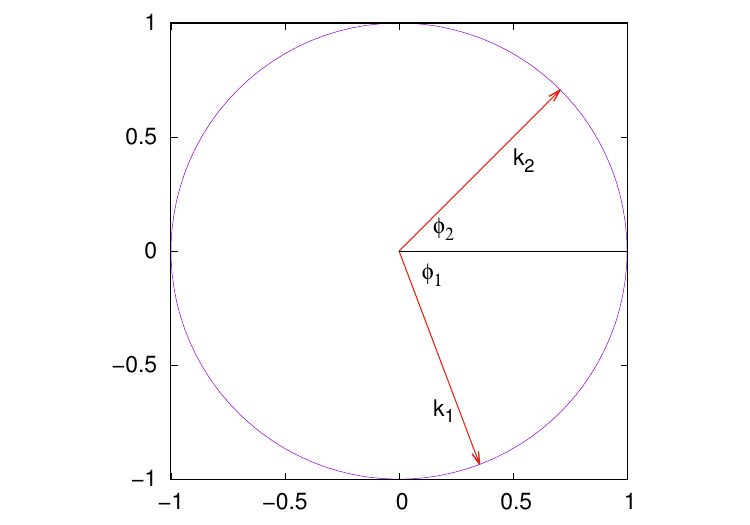}
    \caption{Model circular Fermi surface of radius $k_{F}=1$}
\end{figure}
As we have $|\mathbf{k}|^{2} = 1 $, it is convenient to rewrite the BCS equation in the angle space. The order parameter now depends the angle variable $\varphi, 
 \varphi^{\prime}$:
\begin{align}
\lambda\Delta^{S}(\varphi) &=-\frac{3N}{2\pi}\int_{0}^{2\pi}V(\varphi, \varphi^{\prime})\Delta^{S}(\varphi^{\prime})d\varphi^\prime
\label{eq:singlet}\\
\lambda\Delta^{T}(\varphi) &=\frac{N}{2\pi}\int_{0}^{2\pi}V(\varphi, \varphi^{\prime})\Delta^{T}(\varphi^{\prime})\cos(\varphi-\varphi^{\prime})d\varphi^\prime
\label{eq:triplet}
\end{align}
where $\cos(\varphi - \varphi^{\prime}) = \mathbf{k \cdot k'}$. It is convenient to expand the solution for $\Delta(\varphi)$ in circular harmonics $y_l = \exp(i l \varphi)$, where $\varphi$ is the angular coordinate on the Fermi circle. The inversion constraint $\Delta_{\mathbf{k}} = \Delta_{\mathbf{-k} }$ for both singlet and triplet pairing is satisfied by $l=2n$ so that:
\begin{equation}
\begin{aligned}
    \exp[{2i n (\varphi +k\pi)}]
    &= \exp[{i  (2n\varphi +2nk\pi)}]\\
    &= \exp({2i n \varphi})
\end{aligned}
\end{equation}
with k is an integer and $\varphi + k\pi$ irepresent the inversion of the angle $\varphi$. After $\Delta_{\mathbf{k}}$ is expanded in $y_l$, the problem is reduced to calculating $\lambda$ by direct integration and selecting the maximal value among all even $l=2n$. Note that in the order parameter equation \ref{eq:triplet} for triplet states, there is another dependence on $\varphi$ from $\cos(\varphi-\varphi')$, leading to  $\Delta_{\mathbf{k}}^{T}(\mathbf{k}\cdot\mathbf{k'})$ behaving as  $l=2n+1$, an odd orbital number. 
\section*{Solutions}
\subsection*{Dirac-$\delta$ Distribution}
Introducing the angle $\alpha = 2 \sin^{-1}(Q/2)$ or $Q=2\sin(\alpha/2)$, we then have:
\begin{equation*}
    \begin{aligned}
        |\mathbf{k-k}^{\prime}|-Q &= |\sqrt{2-2\cos(\varphi - \varphi^{\prime})}|-2\sin(\alpha/2)\\
        &= |2\sin(\Tilde{\varphi}/2)|-2\sin(\alpha/2)
    \end{aligned}
\end{equation*}
where $\Tilde{\varphi} =  \varphi^{\prime}-\varphi$. In the simplest case, $f(|\mathbf{k-k}^{\prime}|) = \delta(|\mathbf{k-k}^{\prime}| - Q)$ can be converted into angle variables as $f[|2\sin(\Tilde{\varphi}/2)|-2\sin(\alpha/2)]$. A useful formula is:
\begin{equation*}
\delta(F(x)-a)=\sum_{i}\frac{\delta(x-x_{i})}{|dF/dx|_{x=x_{i}}}
\end{equation*}
in this equation, $i$ labels all solutions of the eq. $F(x)-a=0$ (in our case, there are
two, $\Tilde{\varphi}=\pm\alpha).$ Differentiate $f(\varphi,\varphi^\prime)$ with respect to $\varphi^{'}$ gives $\cos(\Tilde{\varphi}/2)_{\Tilde{\varphi}=\pm\alpha}$. The pairing interaction is then:
\begin{equation}
\begin{aligned}
    V(\varphi,\varphi^\prime)
&= V\left(\frac{\delta({\Tilde{\varphi} -}\alpha)}{\cos(\alpha/{2})} + \frac{\delta({\Tilde{\varphi}+}\alpha)}{\cos(-\alpha/{2})}\right)\\
\label{eq:delta-V}
\end{aligned}
\end{equation}
\paragraph*{Singlet State:}
 With the solution for $\Delta_{\mathbf{k}}^{S}$ from Eq.~\ref{eq:singlet}, the order parameter reads:
\begin{equation*}
    \lambda_n^{S}\exp(2in\varphi)  =-\frac{3N}{2\pi}\int_{0}^{2\pi}V(\varphi\mathbf{,}\varphi^{\prime})\exp(2in\varphi^{\prime})d\varphi^\prime
\end{equation*}
as the integral in the right-hand side (RHS) evaluated with respect to $\varphi'$, we can switch variable $\Tilde{\varphi}$ by dividing both sides with $\exp(2in\varphi)$:
\begin{equation}
    \lambda_n^{S}  =-\frac{3N}{2\pi}\int_{0}^{2\pi}V(\varphi\mathbf{,}\varphi^{\prime})\exp(2in\Tilde{\varphi})d(\Tilde{\varphi})    
\end{equation}
the integral can be solved analytically with the pairing interaction modified by Dirac-$\delta$ distribution Eq.~\ref{eq:delta-V}:
\begin{equation}
\begin{aligned}
    \lambda_n^{S}&=-\frac{3NV}{\pi\sqrt{1-Q^2/4}} T_{2n}(1-Q^2/2)
\end{aligned}
\end{equation}
where we expressed the solution for $\lambda_{n}^{S}$ with Chebyshev polynomials of the first kind $T_{n}(\cos\theta) = \cos(n\theta)$. Another way to rewrite this expression more compactly is to introduce an auxiliary variable $\tilde Q=\cos(\alpha/2)=\sqrt{1-Q^2/4}$. Then 
\begin{equation}
\lambda_n^{S}=-\frac{3NV}{\pi\tilde Q}T_{2n}(2\tilde Q^2-1)
\end{equation}

\paragraph*{Triplet State:}
Following analogous steps, we derive a corresponding order parameter equation with Eq. \ref{eq:triplet} for triplet states:
\begin{equation}
    \lambda_n^{T}  =\frac{N}{2\pi}\int_{0}^{2\pi}V(\varphi\mathbf{,}\varphi^{\prime})\exp(2in\Tilde{\varphi})
    \cos(\Tilde{\varphi})
    d(\Tilde{\varphi})    
\end{equation}
in which the pairing strength constant $\lambda_n^{T}$ can be derived analytically in a manner similar to that of the singlet case: 
\begin{eqnarray}
    \lambda_n^{T} 
    &=&\frac{NV(2\tilde Q^2-1)}{\pi\tilde Q} T_{2n}(2\tilde Q^2-1)
\end{eqnarray}

We then create a diagram to determine, within the range of $\alpha$ from 0 to $\pi$, which state yields the maximum $\lambda_{\text{max}}$ and find the corresponding orbital number: $l=2n$ for the singlet state (s, d, g) and $l=2n+1$ for the triplet state (p, f, h), with $n$ in the range [0, 2], respectively. 

\subsection*{Gaussian Distribution}
Following analogous steps as with the Dirac-$\delta$ distribution, the pairing interaction $V(|\mathbf{k-k'}|)$ for the Gaussian function can be expressed in terms of the angle variable $\Tilde{\varphi}$:
\begin{equation}
    V(\Tilde{\varphi}) =
    \frac{V}{\kappa \sqrt{\pi}} \exp\left\{-\frac{[2|\sin(\tilde \varphi/2)| - 2 \sin(\alpha/2)]^2}{\kappa^2}\right\}
\end{equation}
\paragraph*{Singlet State:} The pairing strength constant equation given the Gaussian model for pairing interaction in the variable $\Tilde{\varphi}$ is as follows:
\begin{equation}
\begin{aligned}
\lambda^{S}_{n} =&  -\frac{3NV}{2\kappa\pi^{3/2}}\int_{0}^{2\pi} \exp(2in \tilde \varphi) \\
& \exp\left\{-\frac{[2|\sin(\tilde \varphi/2)| - 2 \sin(\alpha/2)]^2}{\kappa^2}\right\}d\tilde \varphi
\end{aligned}
\label{eq:GH_S,og}
\end{equation}
note that the integrand $F(\tilde\varphi)$ under the integral in the RHS of Eq.~\ref{eq:GH_S,og} is an even function, meaning that
$F(\tilde\varphi) = F(\tilde\varphi+\pi)$. That leads to
\begin{align*}
\int_{0}^{2\pi} F(\tilde\varphi) d\tilde \varphi &= \int_{0}^{\pi} F(\tilde\varphi) d\tilde \varphi + \int_{\pi}^{2\pi} F(\tilde\varphi) d\tilde \varphi \\
&= 2\int_{0}^{\pi} F(\tilde\varphi) d\tilde \varphi
\end{align*}
which simplifies Eq.~\ref{eq:GH_S,og} to:
\begin{equation*}
\begin{aligned}
\lambda^{S}_{n} =&  -\frac{3NV}{\kappa\pi^{3/2}}\int_{0}^{\pi} \cos(2n \tilde \varphi) \\
&\exp\left\{-\frac{[2\sin(\tilde \varphi/2) - 2 \sin(\alpha/2)]^2}{\kappa^2}\right\}d\tilde \varphi
\end{aligned}
\end{equation*}
it is more favorable to solve Eq.~\ref{eq:GH_S,og} numerically due to its complexity. For large values of $\kappa$, the equation can be solved straightforwardly by Simpson's rule. 
As $\kappa$ approaches small values, the expression under the integral  Eq.~\ref{eq:GH_S,og} varies more and more rapidly, making Simpson's rule impractical. A useful numerical method for small $\kappa$ is Gauss-Hermite quadrature. Introducing 
\begin{equation*}
    x = \frac{2|\sin(\tilde \varphi/2)| - 2 \sin(\alpha/2)}{\kappa}
\end{equation*}
Eq.~\ref{eq:GH_S,og} is then expressed in terms of the variable $x$ as follows:
\begin{equation}
\begin{aligned}
\lambda^{S}_{n} \approx  &-\frac{3NV}{\pi^{3/2}}
\int_{-\infty}^{\infty} \exp(-x^{2})\\
&\frac{\cos\{4n\arcsin[x\kappa/2 +\sin(\alpha/2)]\}}{\sqrt{1-[x\kappa/2 +\sin(\alpha/2)]^2}} dx
\end{aligned}
\label{eq:GH_S,trans}
\end{equation}
rapidly converging the regime of small $\kappa$, where Simpson's rule fails. Applying Gauss-Hermite quadrature, we have:
\begin{equation*}
    \lambda_n^S \approx \frac{-3NV}{\pi^{3/2}} \sum_{i=1}^{n} w_{i}\frac{\cos\{4n\arcsin[x_i\kappa/2 +\sin(\alpha/2)]\}}{\sqrt{1-[x_i\kappa/2 +\sin(\alpha/2)]^2}}
\end{equation*}
\paragraph*{Triplet State:} Analogously to the singlet case, the pairing strength constant of triplet state equation considering the Gaussian model for pairing interaction in the variable $\Tilde{\varphi}$ is as follows:
\begin{equation}
\begin{aligned}
\lambda^{T}_{n} =& \frac{NV}{2\kappa\pi^{3/2}}\int_{0}^{2\pi} \exp(2in \tilde \varphi) \cos(\tilde\varphi) \\
& \exp\left\{-\frac{[2|\sin(\tilde \varphi/2)| - 2 \sin(\alpha/2)]^2}{\kappa^2}\right\}d\tilde \varphi\\
\end{aligned}
\label{eq:GH_T,og}
\end{equation}
with integrand being an even function, Eq.~\ref{eq:GH_T,og} can be transformed into:
\begin{equation*}
    \begin{aligned}
        \lambda^{T}_{n} =& \frac{NV}{\kappa\pi^{3/2}}\int_{0}^{\pi} \cos(2n \tilde \varphi) \cos(\Tilde{\varphi}) \\& 
\exp\left\{-\frac{[2\sin(\tilde \varphi/2) - 2 \sin(\alpha/2)]^2}{\kappa^2}\right\} d\tilde \varphi
    \end{aligned}
\end{equation*}

Implementing a similar procedure as in the singlet case, we can analyze the solution for large and small values of \(\kappa\):

\begin{itemize}
 \item{Values in the upper range of {${\kappa}$}}: 
 Eq.~\ref{eq:GH_T,og} can be solved numerically using Simpson’s rule. 
 \end{itemize}
 
\begin{itemize}
 \item{Values in the lower range of {${\kappa}$}}: The linearized order parameter reads
 \begin{equation}
\begin{aligned}
\lambda^{T}_{n} \approx  &\frac{NV}{\pi^{3/2}}
\int_{-\infty}^{\infty} \cos\{2 \arcsin[x\kappa/2 + \sin(\alpha/2)]\}\\
&\frac{\cos\{4n\arcsin[x\kappa/2 +\sin(\alpha/2)]\}}{\sqrt{1-[x\kappa/2 +\sin(\alpha/2)]^2}} \exp(-x^{2}) dx
\end{aligned}
\label{eq:GH_T,trans}
\end{equation}
 \end{itemize}

Gauss-Hermite quadrature is an appropriate method for obtaining a valid approximation:
\begin{equation*}
\begin{aligned}
    \lambda_n^T \approx  \frac{NV}{\pi^{3/2}} \sum_{i=1}^{n}w_i  &\frac{\cos\{4na \arcsin[x_i\kappa/2 + \sin(\alpha/2)]\}}{\sqrt{1-[x_i\kappa/2+\sin(\alpha/2)]^{2}}}\\ 
    &\cos\{2 \arcsin[x_i\kappa/2 + \sin(\alpha/2)]\}
\end{aligned}
\end{equation*}

\section*{RESULTS}
\subsection*{Dirac-$\delta$ Distribution}
\begin{figure}[H]
\centering\includegraphics[width=0.95\linewidth]{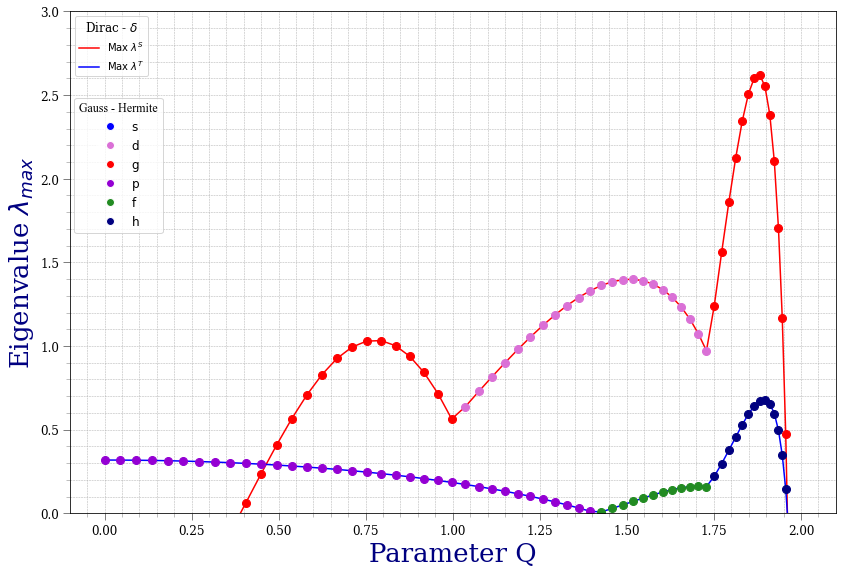}
\caption{Phase Transitions of Singlet and Triplet States modeled by Dirac-$\delta$ function and Gaussian functions with $\kappa = 0.001$}
\label{fig:Dirac-GH}
\end{figure}

The eigenvalue interaction parameter $\lambda_{max}$ is intrinsically positive, aligned with physical expectations of interaction strengths within the system; hence, the diagram excludes values of negative eigenvalues. The red line, representing the singlet pairing interaction shows dominance over the triplet pairing interaction (blue line), except for a small range of $0<|\mathbf{Q}|<0.45$. Within this interval, p-waves states show a relative increase in triplet interaction strength. In the dominant domain of singlet states, there are mostly g-wave states and d-wave states.

    The Gaussian distribution closely approximates the Dirac-$\delta$ model as $\kappa$ approach zero, as illustrated in the Fig.~\ref{fig:Dirac-GH} above, where we compare phase transition of the Gaussian model with width $\kappa = 0.001$ over the Dirac-$\delta$ model. The result confirmed the model's hypothesis.
    
\subsection*{Gaussian Distribution}

\begin{figure}[ht]
\centering
\includegraphics[width=0.95\linewidth]{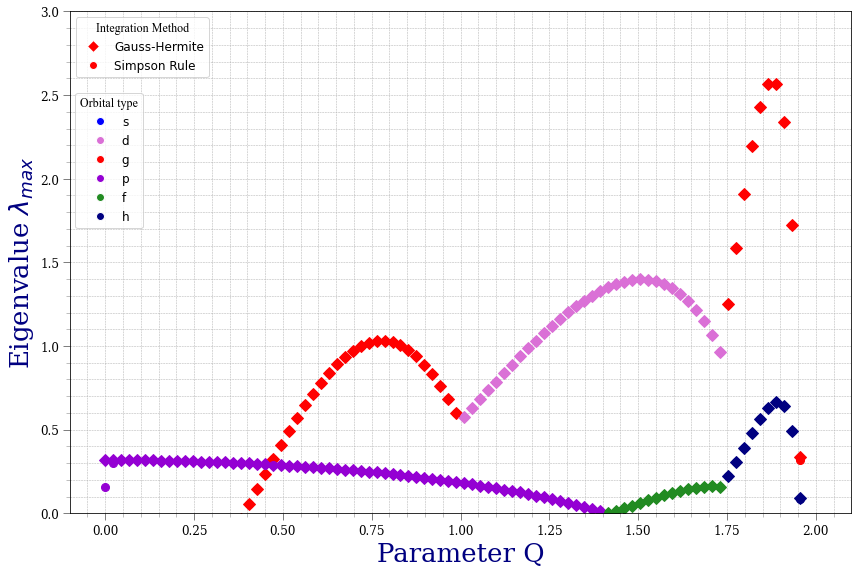}
\caption{Phase Transitions of Singlet and Triplet States modeled by Gaussian function with $\kappa = 0.02$ computed by Gauss-Hermite quadrature vs Simpson's rule}
\label{fig:GH-SR}
\end{figure}

    The Gaussian model represents a more realistic approximation compared to the idealized Dirac-$\delta$ model. By varying the width parameter away from zero, we can develop a comprehensive phase diagram that distinguishes between singlet and triplet states.

    
    As discussed, employing Simpson's rule becomes more appropriate for solving Eqs.~\ref{eq:GH_S,og} and \ref{eq:GH_T,og} over a broader range of $\kappa$, with a particular focus on values starting from $\kappa = 0.02$. As illustrated in Fig.~\ref{fig:GH-SR}, the physical significance of phase dominance is preserved in both cases when $\kappa = 0.02$. This method enables the construction of a detailed phase diagram that examines the interaction parameter $\lambda$, with a focus on identifying the values of $n$ that optimize $\lambda$. This, in turn, provides deeper insights into the system's behavior across different parameter regimes.
\begin{figure}[H]
\begin{minipage}{0.93\linewidth}
\centering
\includegraphics[width=\linewidth]{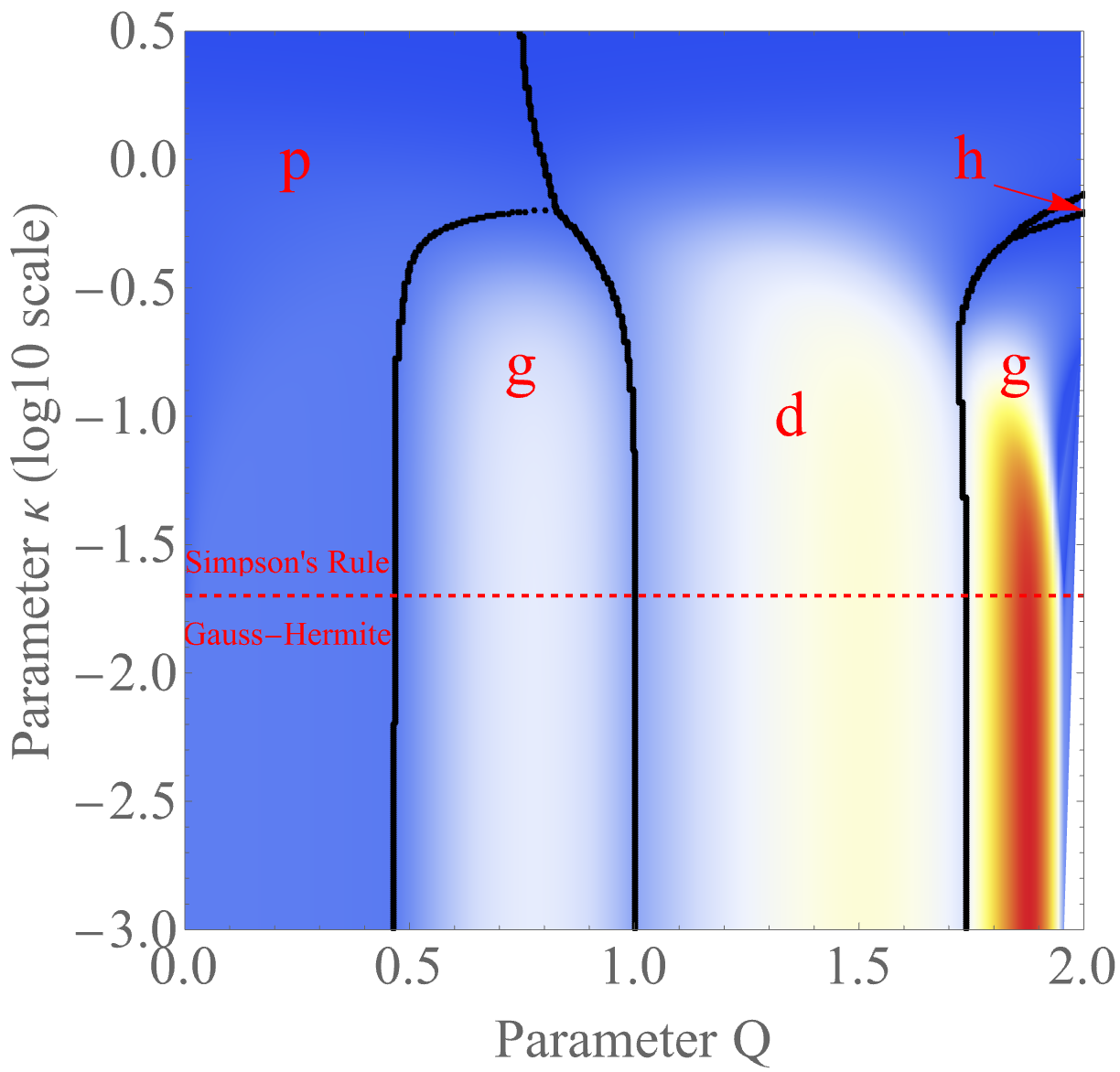}
\end{minipage}%
\begin{minipage}{0.07\linewidth}
\centering
\includegraphics[width=\linewidth]{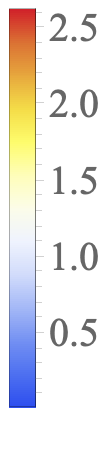}
\end{minipage}
\caption{Phase diagram as a function of the spin-fluctuation wave vector $Q$ and the fluctuation spectrum width $\kappa$. The color corresponds to the absolute values of the largest eigenvalue 
$\lambda_{\max}$ 
of the interaction matrix, and the diagram is cut at $\lambda_{\max}>0.01$. 
The symmetry of the corresponding superconducting state is marked for the corresponding stability regions.}
\label{fig:phase_diagram}
\end{figure}
The phase diagram presented, Fig.~\ref{fig:phase_diagram}, demonstrates the instability of the dominant state as it varies with parameters $\kappa$ and $Q$ in the ranges $[10^{-3}, 10^{1}]$ and $[0, 2]$, respectively. Singlet-wave states are not observed at low Q in the ranges $[0, 0.45]$; only triplet-wave states present, with the prevalence of p-wave states. Nevertheless, due to the factor of 3 associated with the rotational invariance of singlet-wave pairing, singlet-wave states generally dominate the interaction strength (g-wave and d-wave). For $\kappa$ deviates from 0, simulating Dirac-$\delta$ distribution, and \( Q \) approaches its maximum value of 2, there is an absence of pairing interactions for any orbitals, reflecting the physical interpretation the unpopularity  of the spin-fluctuation pairing for antiferromagnetic pairs of electron. A thorough analysis of the system's behavior can be obtained through diagrams of orbital gap functions of each wave state.

\begin{table}[H]
\centering

\begin{tabular}{|c|c|c|c|}
\hline
  &S & {L} & $\Delta(\mathbf{k})$\text{or} $\textbf{d}(\mathbf{k})$ \\
\hline
S&0 & s  & 1 \\
T&1 & p  & $\cos \phi$ \\
S& 0 & d  & $\cos 2\phi$ \\
T& 1 & f  & $\cos 3\phi$\\ 
S& 0 & g  & $\cos 4\phi$ \\
T& 1 & h  & $\cos 5\phi$ \\
\hline
\end{tabular}
\caption{Symmetry of Gap Functions of States}
\label{tab:pairing}
\end{table}
    
Table \ref{tab:pairing} shows the symmetry of gap functions for states depicted in Fig.~\ref{fig:phase_diagram}, categorized by orbital symmetry (L) and the commonly used s-, p-, d-wave symmetries in a two-dimensional surface. The basis functions for the scalar singlet (\(S = 0\)) order parameter \(\Delta\), and for the vector triplet (\(S = 1\))  order parameter \(\mathbf{d}\) are listed. 

Note that the gap functions for spin-triplet states with the same angular momentum can have degenerate states. For example, the p-wave symmetry on an isotropic 2D Fermi surface has five degenerate (without spin-orbit coupling) representations:
\begin{align*}
    &k_{x} \hat{x} \pm k_y \hat{y},\\
    &k_{y} \hat{x} \pm k_x \hat{y},\\
    &k_{x} \hat{z} + i k_y \hat{z} \quad (\text{Anderson-Brinkmann-Morel state}),
\end{align*}
all of which can be expressed by \( k_{x} \hat{x} + k_y \hat{y}\), the isotropic Balian-Werthamer state. Hence, the gap functions listed in Table \ref{tab:pairing} effectively represent all these degenerate states.

\begin{figure}[H]
\centering
\includegraphics[width=0.44\linewidth]{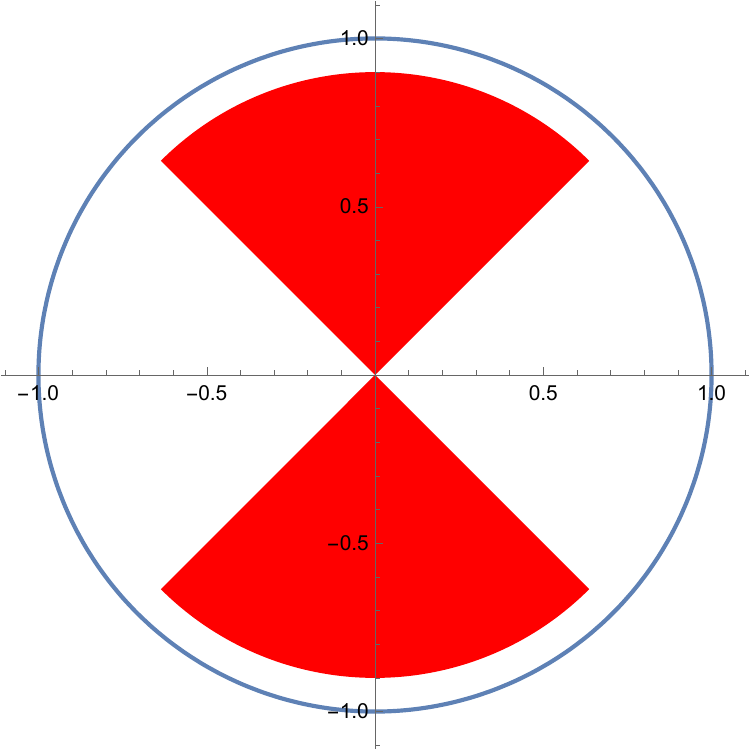}
\centering
\includegraphics[width=0.44\linewidth]{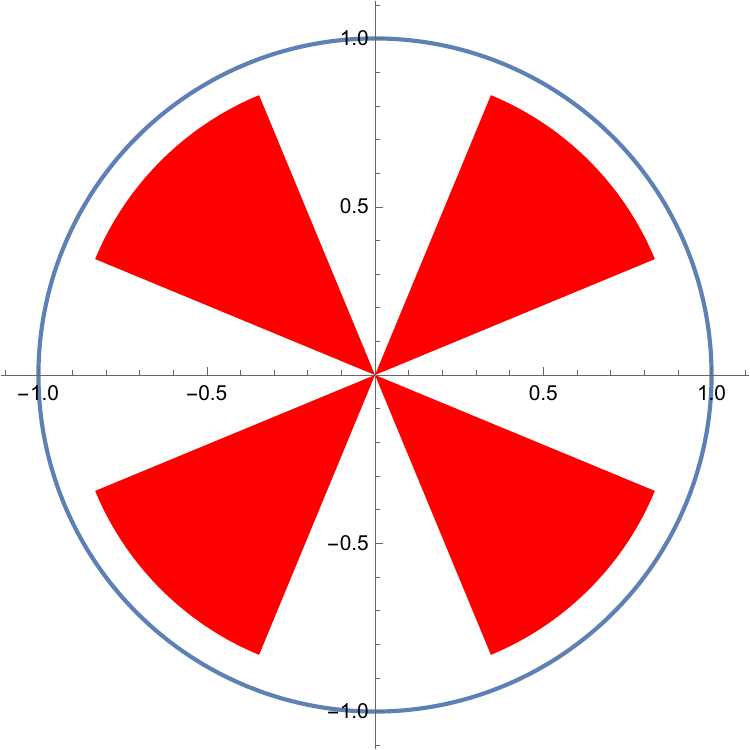}
\caption{Pairing (red) and pair-breaking (white) regions for scattering from the leftmost point on the Fermi surface. Left: d-wave; right: g-wave.}
\label{fig:dg-pairing}
\end{figure}

\begin{figure}[H]
\centering
\includegraphics[width=0.44\linewidth]{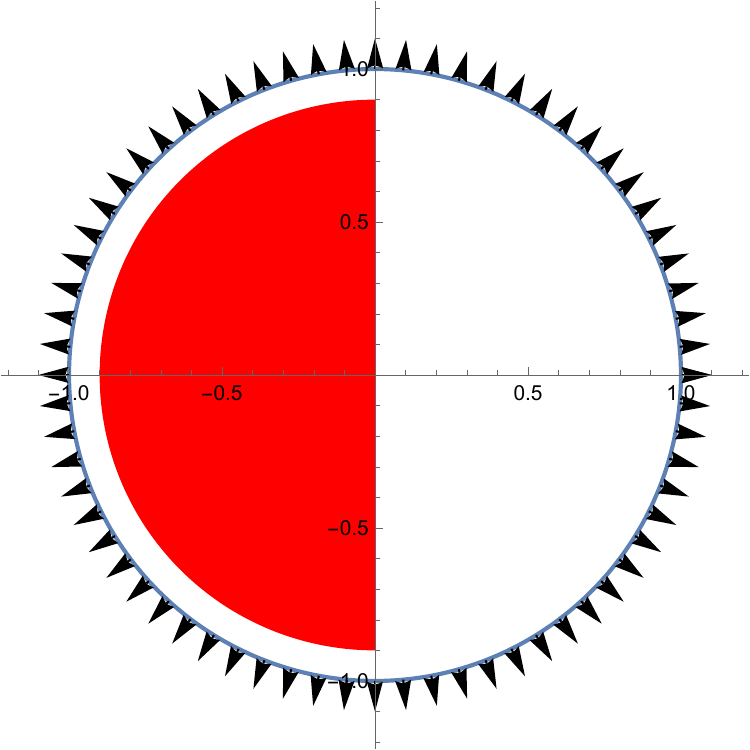}
\includegraphics[width=0.44\linewidth]{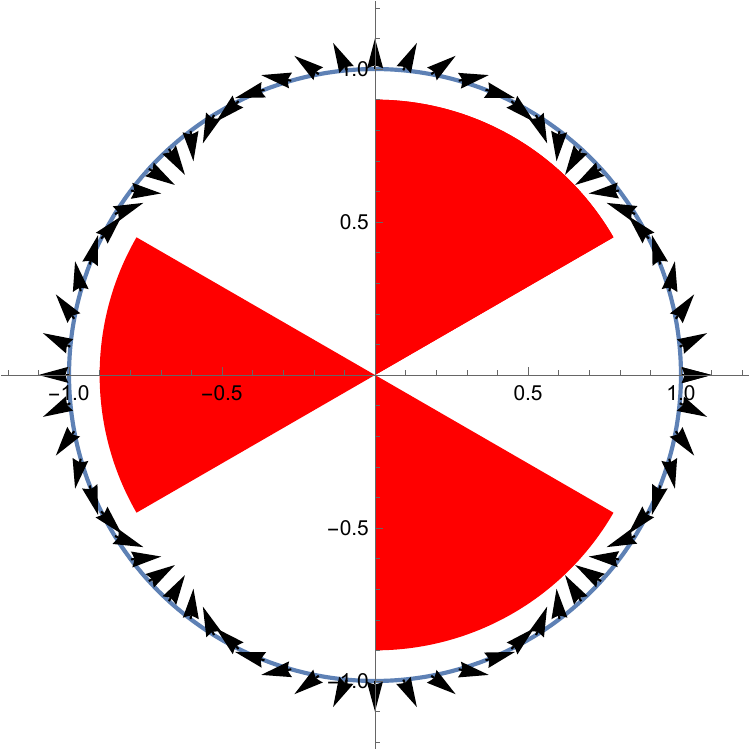}
\centering
\includegraphics[width=0.44\linewidth]{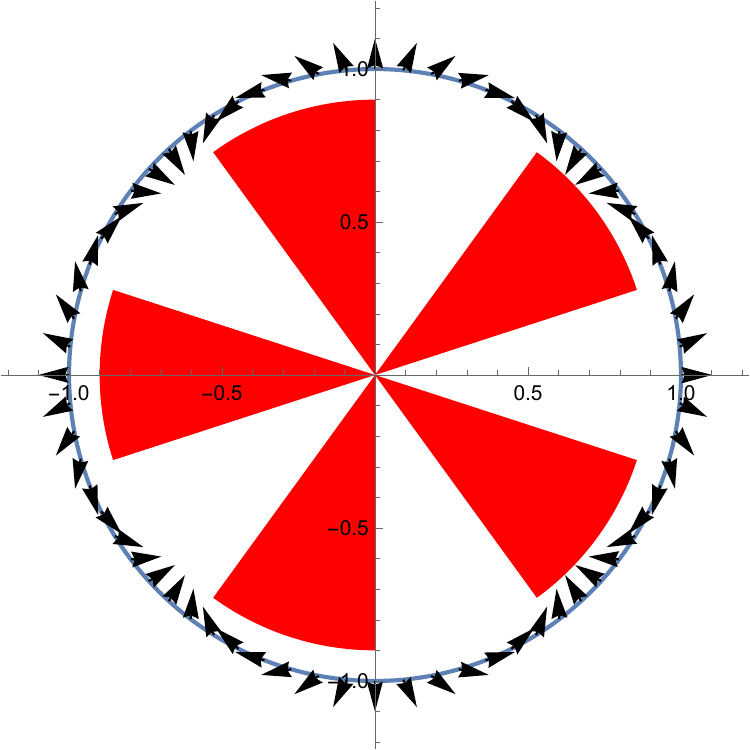}
\caption{Pairing (red) and pair-breaking (white) regions regions for scattering from the leftmost point on the Fermi surface. Left: p-wave; right: f-wave; bottom: h-wave. The direction of the $\mathbf{d}$ is indicated by arrows.}
\label{fig:pfh-pairing}
\end{figure}
 Figs.~\ref{fig:dg-pairing} and \ref{fig:pfh-pairing} provide visual clues to the structure of the phase diagram (Fig.~\ref{fig:phase_diagram}). 
 First of all, at small \(Q\) (i.e., for approximately ferromagnetic spin fluctuations), singlet states cannot for; only triplet-wave states can take advantage of such spin fluctuations. For the p-wave states fluctuations with small, but finite $Q$ are also pairing, therefore it is not very sensitive to the width of the fluctuation spectrum. Indeed we find $p$-wave to be stable at $Q$ up to nearly 0.5, and its range of stability increase with $\kappa$.

 At $\varphi\sim \pi/4$ ($Q=1\sin{\pi/8}\sim 0.76$ the state that is most favored is $g$, albeit its advantage gradually deteriorates when $\kappa$ becomes comparable with $Q$. Not that factor of three in Eq. \ref{d2} additionally favors singlet states. Indeed we see that the critical temperature for the $g$ state is maximized at $Q\approx 0.76$. 

 As $Q$ increases further towards $\sqrt{2}\approx 1.4$ the situation replicates that in the high-Tc cuprates, since this wave vector corresponds to the nearest neighbor antiferomagnetic coupling on a square lattice. Of course, the $d$ state, which corresponds specifically to the $d_{x^2-y^2}$ on the square CuO$_2$ lattices, fits this $Q$ perfectly, and we see another singlet ($d$) maximum at this vector. It is stronger than that for the $g$ wave, because the pairing region is broader. This fact is also responsible for enlargement of the $d$  stability region at larger $\kappa$. 

 Finally, the g state is again becoming well paired for $Q\sim 2\sin{3\pi/8}\approx 1.85$, albeit at slightly larger $Q\sim 2\sqrt{7\pi/16}\approx 1.96$ the h states becoves competitive --- but still loses, because of the factor of three, except in a thin sliver of the phase diagram at $\kappa\approx 0.632$.

As \(Q\) approaches its maximum value of 2, the calculated critical temperature drop precipitously. This can be rationalized as follows: spin fluctuations with $Q=2$ are attractive for triplet pairing, but, since $\mathbf{k\cdot k}=-1$ for the two  opposite points, it is pair-breaking. For the singlet pairing it is repulsive, but then the sign of the order parameter must be the same for those point, by symmetry, so it is again pair-breaking. Larger widths slightly alleviate this frustration, so that pairing with, consequently, $g$, $h$ and $d$ symemtry becomes possible --- but very weak.

\section*{Conclusions} 

Spin-fluctuation pairing interactions with varying momentum-space structures can lead to a rich variety of unconventional pairing states, many of which exhibit unique symmetries and parities that are unattainable through individual interactions alone. Even the extremely simple toy model of an isotropic 2D Fermi surface with isotropic spin fluctuations can lead, as a functions of the position and and width of the spin fluctuation maximum, to a surprisingly rich phase diagram with singlet and triplet pairing states with angular momenta $L$=1, 2, 3, 4 or 5. While some of our findings may be model specific, such as a rather tiny stability region of the $h$-wave state, most of them are quite generic: dominance of the $p$-wave paring close to ferromagnetism, $d$-pairing in the regime similar to nearest-neighbor spin correlations in cuprates, and strong stability of $g$-wave for other spin fluctuation wave vector. Another generic finding is the increased stability of $p$- and $d$-waves at the expense of the $g$-wave at larger-width fluctuation spectra.
 
\section*{Acknowledgments}
We gratefully acknowledge financial support from the Intensive Undergraduate Research Scholars Program (URSP) at the Office of Student Scholarship, Creative Activities, and Research (OSCAR), George Mason University. The project was also partially supported by  the National Science Foundation (Award No. 2214194)
        and through resources provided by
        the Office of Research Computing at George Mason University
        (\url{https://orc.gmu.edu}) funded in part
        by grants from the National Science Foundation
        (Award Nos.\ 1625039 and 2018631). We also extend our appreciation to Dr. Rafael Fernandes who suggested to us the idea of this project.

\bibliography{bib}

\end{document}